\documentclass[conference]{IEEEtran}
\IEEEoverridecommandlockouts

\ifCLASSINFOpdf
\else
\fi
\usepackage{xcolor,soul,framed} 
\colorlet{shadecolor}{yellow}
\usepackage[pdftex]{graphicx}
\graphicspath{{../pdf/}{../jpeg/}}
\DeclareGraphicsExtensions{.pdf,.jpeg,.png}
\usepackage{cite}

\usepackage[ruled,vlined]{algorithm2e}
\usepackage{algorithmic}
\makeatother
\makeatother
\usepackage{amsmath,amssymb}
\SetKw{Break}{break out of the loop}
\usepackage{graphicx}
\usepackage{textcomp}
\usepackage{comment}
\usepackage{units}
\usepackage{url}
\usepackage{geometry}
\usepackage{xcolor}
\definecolor{myblue}{rgb}{0.0, 0.5, 1.0}
\definecolor{myred}{rgb}{1.0, 0.13, 0.32}
\definecolor{mygreen}{rgb}{0.31, 0.68, 0.07}
\usepackage[pdftex]{graphicx}
\DeclareGraphicsExtensions{.pdf,.jpeg,.png}
\def\BibTeX{{\rm B\kern-.05em{\sc i\kern-.025em b}\kern-.08em
    T\kern-.1667em\lower.7ex\hbox{E}\kern-.125emX}}

\usepackage{pgfplots}
\usepackage{pgfplotstable}
\usepackage{tikz}
\usetikzlibrary{calc}

\usepackage{balance}
\begin{document}

\title{ {\huge Low-Complexity Linear Programming Based Decoding of Quantum LDPC codes}
\thanks{The work of S. Javed and M. F. Flanagan was supported by Science Foundation Ireland under the US-Ireland R\&D Partnership Programme (CoQREATE, Grant Number SFI/21/US-C2C/3750). The work of F. Garcia-Herrero was supported by the QuantERA grant EQUIP (Spain MCIN/AEI/10.13039/501100011033, grant PCI2022-132922), funded by Agencia Estatal de Investigación, Ministerio de Ciencia e Innovaci\'{o}n, Gobierno de España and by the European Union ``NextGenerationEU/PRTR". Bane Vasi\'{c} acknowledges the support of the NSF under grants CIF-1855879, CCF-2100013, CIF-2106189, CCSS-2027844, CCSS-2052751, CIF-2106189, and ERC-1941583 (CoQREATE), as well as the support of NASA through the SURP Program. Bane Vasi\'{c} has disclosed an outside interest in Codelucida to the University of Arizona. Conflicts of interest resulting from this interest are being managed by The University of Arizona in accordance with its policies.} \vspace{-0.4cm}
}
\newgeometry {top=25.4mm,left=19.1mm, right= 19.1mm,bottom =23mm}
\author{Sana Javed$^{*}$, Francisco Garcia-Herrero$^{\dagger}$, Bane Vasi\'{c} $^{\ddagger}$ and Mark F. Flanagan$^{*}$ \\
$^{*}$ School of Electrical and Electronic Engineering, University College Dublin, Belfield, Dublin 4, Ireland, \\
$^{\dagger}$ Department of Computer Architecture and Automatics, Complutense University of Madrid, Madrid, Spain\\
$^{\ddagger}$ Department of Electrical and Computer Engineering, The University of Arizona, Tucson, AZ, 85721 USA\\
Email: sana.javed@ucdconnect.ie, francg18@ucm.es, vasic@ece.arizona.edu, mark.flanagan@ieee.org
\vspace{-0.4cm}}
\maketitle
\begin{abstract}
This paper proposes two approaches for reducing the impact of the error floor phenomenon when decoding quantum low-density parity-check codes with belief propagation based algorithms. First, a low-complexity syndrome-based linear programming (SB-LP) decoding algorithm is proposed, and second, the proposed SB-LP is applied as a post-processing step after syndrome-based min-sum (SB-MS) decoding. For the latter case, a new early stopping criterion is introduced to decide when to activate the SB-LP algorithm, avoiding executing a predefined maximum number of iterations for the SB-MS decoder. Simulation results show, for a sample hypergraph code, that the proposed decoder can lower the error floor by two to three orders of magnitude compared to SB-MS for the same total number of decoding iterations.

\end{abstract}


\maketitle

\IEEEdisplaynontitleabstractindextext

\IEEEpeerreviewmaketitle

\section{INTRODUCTION}

Quantum algorithms that involve a large number of qubits (between thousands and millions) play a pivotal role in advancing quantum computing, unlocking the ability to tackle intricate problems that are beyond the reach of classical computers \cite{arute2019quantum}. However, noise in quantum systems, generated by factors such as decoherence, crosstalk, and environmental interference, represents a significant hurdle \cite{calderbank1996}, making it challenging to maintain the integrity of quantum computation, especially when the number of qubits scales \cite{arute2019quantum}--\cite{calderbank1997quantum}. 

To reduce the effects of noise, many new codes and decoders for quantum error correction have been designed over the past few decades \cite{calderbank1996}. Some of the most intensively studied codes are surface codes \cite{wootton_2012}, initially suggested for noisy intermediate-scale quantum (NISQ) devices. Unfortunately, these codes have some limitations when the objective is to protect a large number of logical qubits, as they have a low code rate and consequently an impractical overhead in terms of physical qubits. 

As an alternative with a higher code rate, quantum low-density parity-check (QLDPC) codes offer benefits by reducing the qubit overhead and also improving the minimum distance \cite{babar2015fifteen}\textcolor{black}{, \cite{panteleev2022_asymp}}. Despite their potential, there are still open problems such as the fault-tolerant implementation of the encoding process on the quantum side and the design of efficient decoders that can obtain very low logical error rates \cite{breuckmann2021quantum}. This paper will focus on addressing the latter limitation.



One of the main problems in performing efficient decoding for a QLDPC code is the existence of short cycles within the code's Tanner graph, which result in an error floor when belief propagation (BP) based decoders are applied. This error floor effect means that an improvement in the physical error rate of the qubits does not translate into an improvement in the logical error rate \cite{Raveendran_2021}. To mitigate this problem, two main categories of approach have been followed: the use of a new standalone decoder based on alternative approaches to standard BP, and applying a post-processing technique after BP decoding.



For the first category of approach, some more powerful decoding techniques such as the syndrome-based generalized belief propagation (GBP) algorithm \cite{raveendran2019syndrome} have been proposed to improve the decoding performance. Unfortunately, these require a high computational complexity, so additional simplifications are necessary to obtain practical real-time solutions. Some neural network based decoders for QLDPC codes have also been explored \cite{liu2019neural}. However, training and deploying these algorithms for large QLDPC codes can be computationally intensive and resource-demanding when the number of qubits grows. 


For the second category, ordered statistics decoding (OSD) was \textcolor{black}{proposed in \cite{Panteleev_2021}} as a post-processing technique for BP. By seamlessly integrating OSD when BP decoding fails, this technique can significantly enhance the error correction capability for some specific QLDPC codes, providing improved error rates after post-processing. However, OSD decoders are computationally complex and introduce an extremely large latency due to the requirement, in one of the steps, of performing Gaussian elimination for the inversion of a dynamic submatrix of the code's parity-check matrix; thus, an efficient hardware implementation that can finish decoding within the required time interval (of hundreds of nanoseconds to several microseconds) is not feasible \cite{valls2021syndrome}.

Stabilizer inactivation decoding for QLDPC codes was also recently introduced in \cite{du2022stabilizer} as an alternative post-processing method; this was shown to outperform OSD in lowering the error floor while also reducing the decoding complexity. However, it is a list decoding method (the list being computed according to the BP output), the complexity of which grows with the length of the list. Serial processing of the list can lead to information loss due to the short decoherence time, while parallel processing would require high power consumption. This highlights the importance of continuing the search for alternative decoding techniques.



Besides these previous algorithms, linear programming (LP) based decoding has also been recently explored as a solution for quantum codes. As LP decoding proved to be capable of reducing the error floor for classical LDPC codes \cite{feldman2005using}, some works such as \cite{li2018lp} and \cite{fawzi2021linear} adapted classical LP decoding algorithms to QLDPC codes, also providing theoretical performance guarantees. Nevertheless, these decoders require the solution of linear programs and thus these algorithms are not attractive from a complexity and latency perspective.

In this paper, a low-complexity iterative syndrome-based linear programming (SB-LP) decoder is proposed for decoding QLDPC codes, which can be used as a standalone decoding algorithm or as a post-processing step after syndrome-based min-sum (SB-MS) decoding to improve the behavior in the error floor region and the threshold of the code. Unlike the standalone SB-MS decoder, which exhibits a high error floor for several QLDPC code classes, the proposed combination of SB-MS with SB-LP improves the error floor with lower complexity \textcolor{black}{than OSD and without the need for list decoding as required by stabilizer inactivation.} 
We also propose a new early stopping criterion for the SB-MS decoder based on the syndrome Hamming distance, using which the algorithm can automatically detect misconvergence of the SB-MS decoder and use this to trigger the SB-LP post-processing. 

\section{Background}
This section summarizes the depolarizing error model used in our work, and provides a brief description of the SB-MS decoder which is a \textit{de facto} standard for low-complexity BP based decoding. In the following, all vectors are assumed to be row vectors.


\subsection{Quantum Noise Model}


Quantum gates, crosstalk, and other processes applied to qubits (including maintaining them in an idle state) result in bit-flip errors, phase-flip errors, or a combination of both. The behavior of errors in a quantum processor can be described by using a depolarizing noise model \cite{nielsen2010quantum}. Errors resulting from bit-flips are described by Pauli $\textrm{X}$ operators and have a probability of occurrence denoted by $p_\textrm{X}$. Phase-flip errors are described by Pauli $\textrm{Z}$ operators and have a probability of occurrence $p_\textrm{Z}$. Errors involving both bit-flips and phase-flips are described by Pauli $\textrm{Y}$ operators and occur with probability $p_\textrm{Y}$. We consider an i.i.d. symmetric depolarizing error model for the qubits, such that each qubit experiences an independent depolarizing error probability $p$ and $p_\textrm{X}=p_\textrm{Y}=p_\textrm{Z}=p/3$. Bit-flip and phase-flip errors are represented by binary error vectors $\textbf{e}_\textrm{X}$ and $\textbf{e}_\textrm{Z}$, respectively, in which the $j$-th entry is equal to $1$ in the presence of a corresponding error on the $j$-th qubit, and is equal to $0$ if there is no such error. An $[[n,k]]$ QLDPC code, where $k$ logical qubits are encoded using $n$ physical qubits, can be characterized by two $m \times n$ binary parity-check matrices $\textbf{H}_\textrm{X}$ and $\textbf{H}_\textrm{Z}$; the corresponding length-$m$ binary syndrome vectors $\textbf{s}_\textrm{X}={\textbf{e}_\textrm{X}}{{\textbf{H}_\textrm{Z}}^T}$ and $\textbf{s}_\textrm{Z}={\textbf{e}_\textrm{Z}}{{\textbf{H}_\textrm{X}}^T}$ are measured from the quantum system and can be used to correct $\textrm{X}$ and $\textrm{Z}$ errors, respectively \cite{babar2015fifteen}. 
For ease of exposition, in the rest of this paper we will focus on correction of only one type of error, and we will use the simpler notations $\textbf{H}$, $\textbf{e}$ and $\textbf{s}$ in place of $\textbf{H}_\textrm{Z}$, $\textbf{e}_\textrm{X}$ and $\textbf{s}_\textrm{X}$, respectively. Also, we will assume the syndrome measurement to be error-free.

\subsection{Tanner Graph}

The Tanner graph serves as a graphical representation of the $m \times n$ parity-check matrix $\textbf{H} = (\textrm{H}_{ij})$. This graph consists of $n$ variable nodes (VNs) and $m$ check nodes (CNs), connected by a set of edges $\mathcal{E}$. For each $i\in \{1,2,\ldots,m\}$ and $j\in \{1,2,\ldots,n\}$, the $i$-th CN is connected to the $j$-th VN by an edge (i.e., $(i,j) \in \mathcal{E}$) if and only if ${\textrm{H}_{ij}} = 1$. $\mathcal{N}_i$ denotes the set of neighbors of the ${i}$-th CN, while $\mathcal{N}_j$ denotes the set of neighbors of the $j$-th VN. For ease of exposition, we assume that the Tanner graph is \textit{regular}, i.e., all CNs have the same degree $d_c$ and all VNs have the same degree $d_v$. In the QLDPC decoding context, VNs correspond to entries of the estimated error pattern ${\textbf{e}} = (e_1 \; e_2 \; \cdots \; e_n)$ while CNs correspond to the entries of the syndrome vector $\textbf{s} = (s_1 \; s_2 \; \cdots \; s_m)$. 



\subsection{Syndrome-Based Min-Sum (SB-MS) Algorithm}
The SB-MS algorithm is a low-complexity approximation to BP which replaces the complex CN operation with a simpler minimum operation \cite{fossorier1999reduced}. Its operation is given in Algorithm \ref{algo:SB-MS}.

The SB-MS decoder seeks to find an error pattern $\hat{\textbf{e}}$ that matches the measured syndrome $\textbf{s}$. It does this by iteratively exchanging messages between the VNs and CNs (lines 4 to 7), making hard decisions to produce an estimate of the error vector $\hat{\textbf{e}}$ (line 8), and generating the corresponding syndrome $\hat{\textbf{s}}$ (line 9). The algorithm terminates when either this syndrome matches the measured syndrome or the maximum number of iterations $I^{\textrm{(MS)}}_{\textrm{max}}$ is reached.

Here $u_{i,j}$ represents the message from the $i$-th CN to the $j$-th VN, and $v_{i,j}$ denotes the message sent from the $j$-th VN to the $i$-th CN. The hard decision operation is defined as 
\begin{equation*}
\mathrm{HD}(x)= 
\left\{
\begin{array}{ll}
0 & \text{if } x > 0, \\
1 & \text{otherwise.}
\end{array}
\right.
\end{equation*}
and we also define the function
\begin{equation*}
\mathrm{sgn}(x) = 1 - 2\mathrm{HD}(x)= 
\left\{
\begin{array}{ll}
1 & \text{if } x > 0, \\
-1 & \text{otherwise.}
\end{array}
\right.
\end{equation*}
${\lambda_j} = \log \left( \frac{P(e_j = 0)}{P(e_j = 1)} \right)$ represents the \textit{a priori} log-likelihood ratio (LLR) of $e_j$, while $\alpha$ is a scaling factor that accelerates the convergence of the algorithm. 

\begin{algorithm}
\label{algo:SB-MS}
\caption{Syndrome-Based Min-Sum (SB-MS) Algorithm}
\begin{algorithmic}[1] 
\STATE Input: syndrome $\textbf{s}$
\STATE Initialize parameters:\\
$\lambda_j$ = $\ln \left( \frac{1-(2p/3)}{2p/3} \right)$,   $\forall j\in \{1,2,\ldots,n\}$ 
\\
$u_{i,j}=0, \forall  (i,j) \in \mathcal{E}$\\
$I=0$, $\hat{\textbf{s}}=\mathbf{0}$
\WHILE {$\textbf{s}\neq \hat{\textbf{s}}$ and $I$ $\leq$ $I^{\textrm{(MS)}}_{\textrm{max}}$}
\STATE \indent \textbf{for} all $ (i,j) \in \mathcal{E}$ (in parallel) \textbf{do}
 \[ v_{i,j} = \lambda_j + \alpha\sum_{i'\in \mathcal{N}_j \backslash \{i\}} u_{i',j}\]
 \STATE \indent \textbf{end for}
 \STATE \indent \textbf{for} all $ (i,j) \in \mathcal{E}$ (in parallel) \textbf{do}
      \!\! \[u_{i,j} =\mathrm{sgn}(s_i) \cdot \!\! \prod_{j'\in \mathcal{N}_{i} \backslash \{j\}}^{}\mathrm{sgn}(v_{i,j'} ) \cdot \! \min_{j'\in \mathcal{N}_{i}\backslash \{j\}}^{}\left| v_{i,j'} \right|\]\vspace{-.4cm}
       \STATE \indent \textbf{end for}
\STATE$\hat e_j= \mathrm{HD}(\lambda_j + \alpha\sum_{i\in \mathcal{N}_j } u_{i,j}), \forall j\in \{1,2,\ldots,n\}$
\STATE \[\hat{\textbf{s}}=\hat{\textbf{e}} \textbf{H}^T \]\vspace{-.7cm}
\STATE $I=I+1$
\ENDWHILE
\RETURN $\hat{\textbf{e}}$
\end{algorithmic}
\end{algorithm}

In the case of certain QLDPC codes, the SB-MS decoder is unable to improve the error rate of the qubits even as the depolarizing error probability improves; this results in an error floor. The presence of this error floor can be attributed to the presence of short cycles within the code's Tanner graph which the SB-MS decoding algorithm struggles to handle. These cycles lead to oscillations between different candidate error vectors during the decoding process so that the SB-MS decoder cannot converge to a unique error pattern that matches the syndrome. In the next section, we propose an approach to reduce or eliminate this error floor phenomenon. 

\section{Proposed Method}
%
%
Linear programming based decoding, originally proposed for classical LDPC codes in \cite{feldman2005using}, relies on a linear programming relaxation approach to achieve a high-performance decoding solution with theoretical guarantees on performance. In \cite{5755808}, Vontobel and Koetter proposed low-complexity iterative decoding algorithms that exploited the unique structure of the linear program to be solved (in particular, exploiting the sparsity of $\textbf{H}$). In the following, we propose an LP-based decoder which is based on modifying Algorithm 1 in \cite{5755808} to align with the unique features and requirements of QLDPC codes. 

\subsection{Proposed Syndrome-based Linear Programming (SB-LP) Decoder}


The details of the proposed SB-LP decoder are given in Algorithm \ref{LP}. This algorithm iteratively updates the estimated error vector $\hat{\textbf{e}}$, guided by the measured syndrome $\textbf{s}$, until convergence is achieved or the maximum number of iterations $I^{\textrm{(LP)}}_{\textrm{max}}$ has been reached. Unlike the SB-MS decoder, the SB-LP decoder does not perform VN-to-CN and CN-to-VN updates  alternately; instead, it simultaneously updates a single value $\overline{u}_{i,j}$ for each edge $(i,j) \in \mathcal{E}$ (line 9). This parallel update approach can significantly improve the decoding speed. Also, $\alpha_1$ serves as a scaling factor designed to expedite the algorithm's convergence.

\begin{algorithm}
\label{LP}
\caption{Proposed Syndrome-Based Linear Programming (SB-LP) Algorithm }
\begin{algorithmic}[1] 
\STATE Input: syndrome $\textbf{s}$
\STATE Initialize parameters:\\
$\lambda_j$ = $\ln \left( \frac{1-(2p/3)}{2p/3} \right)$,  $\forall j\in \{1,2,\ldots,n\}$ 
\\ 
$\overline{u}_{i,j}=0, \forall (i,j) \in \mathcal{E}$\\
$I=0$, $\hat{\textbf{s}}=\mathbf{0}$
\WHILE {$\textbf{s}\neq \hat{\textbf{s}}$ and $I$ $\leq$ $I^{\textrm{(LP)}}_{\textrm{max}}$}
\FOR{all $ (i,j) \in \mathcal{E}$ (in parallel)}
    \STATE\[ S_{i,j} = \lambda_j + \sum_{i'\in \mathcal{N}_j \backslash \{i\}} \overline{u}_{i',j}\]\vspace{-.5cm}
\STATE \indent \indent \textbf{if} $s_i = 0$ \textbf{then}
    \[T^{(0)}_{i,j} = \max_{\textbf{b}\in \mathcal{B}_i,b_j=0} \overline{\textbf{u}}_{i,j} \tilde{\textbf{b}}_j^T \]
    \[T^{(1)}_{i,j} = \max_{\textbf{b}\in \mathcal{B}_i,b_j=1} \overline{\textbf{u}}_{i,j} \tilde{\textbf{b}}_j^T \] \vspace{-.4cm}
    \STATE \indent \indent \textbf{else}
     \[ T^{(0)}_{i,j} = \max_{\textbf{c}\in \mathcal{C}_i,c_j=0} \overline{\textbf{u}}_{i,j} \tilde{\textbf{c}}_j^T \]
    \[ T^{(1)}_{i,j} = \max_{\textbf{c}\in \mathcal{C}_i,c_j=1} \overline{\textbf{u}}_{i,j} \tilde{\textbf{c}}_j^T \] 
\STATE \indent \indent \textbf{end if}
    \STATE $\overline{u}_{i,j} = \frac{\alpha_{1}}{2}(T^{(0)}_{i,j}-T^{(1)}_{i,j} - S_{i,j})$\\
    
\ENDFOR\\
\STATE $\hat e_j= \mathrm{HD} (\lambda_j + \sum_{i'\in \mathcal{N}_j} \overline{u}_{i',j}) ,\forall j\in \{1,2,\ldots,n\}$
\STATE\[\hat{\textbf{s}}=\hat{ \textbf{e}} \textbf{H}^T \]\vspace{-.5cm}
\STATE$I=I+1$
\ENDWHILE
\RETURN $\hat { \textbf{e}}$
\end{algorithmic}
\end{algorithm}

In lines 6 and 7, $\mathcal{B}_i$ denotes the single parity-check code of length $d_c$ consisting of all $2^{|d_c|-1}$ binary sequences having even parity, while $\mathcal{C}_i = \{0,1\}^{|d_c|} \backslash \mathcal{B}_i$ represents the set of all $2^{|d_c|-1}$ binary sequences having odd parity (note that these vectors are indexed by the set $\mathcal{N}_i$). Also, $\tilde{\textbf{b}}_j$ and $\tilde{\textbf{c}}_j$ correspond to the vectors $\textbf{b}$ and $\textbf{c}$ with the $j$-th element removed, respectively. Also, we define the check-perspective edge value vector $\overline{\textbf{u}}_{i,j}=(\overline{u}_{i,j'} )_{  j'\in \mathcal{N}_i \backslash \{j\}}$. 

With respect to Algorithm 1 in \cite{5755808}, which was proposed for decoding classical LDPC codes, we have introduced some modifications that enable the algorithm to decode QLDPC codes. First, we have modified the expressions for ${T^{(0)}_{i,j}}$ and ${T^{(1)}_{i,j}}$ in lines 6 and 7, as they depend on the $i$-th syndrome bit $s_i$; in classical coding, we always have $s_i=0$, while in QLDPC coding we may have $s_i=1$ which necessitates performing the corresponding minimization over the complement $\mathcal{C}_i$ of the single parity check code $\mathcal{B}_i$. Second, we have replaced the ``soft minimum" operator in \cite{5755808} with the minimum operator in order to reduce the complexity (this was found to give negligible loss in performance). Finally, we have replaced the successive edge value update rule in \cite{5755808} by a fully parallel update rule, and we have made some other minor simplifications including removing the (unnecessary) dual variables of the LP.


\subsection{SB-LP as Post-Processing after SB-MS Decoding}
%
%
We can further improve the error floor behavior while reducing the overall decoding complexity by combining the SB-MS and SB-LP decoders. Algorithm \ref{MS+LP} describes our proposed methodology for this. 
First, we note that the SB-MS decoder exhibits an oscillatory behavior when it becomes stuck in a trapping set of the QLDPC code \cite{Raveendran_2021}. In such cases, the decoder is not able to identify an error pattern that matches the measured syndrome, and hence the number of unmatched syndrome elements begins to oscillate as it seeks to converge to more than one error vector within the trapping set. If this behavior can be identified automatically by the decoding algorithm (through an appropriately designed early stopping criterion, as described in Subsection III-C), the SB-LP decoder can then be invoked to complete the decoding successfully (as the SB-MS decoder alone will usually not be capable of converging to a solution in such cases). 
 
Moreover, to facilitate faster convergence of the iterative SB-LP decoder, we take advantage of the calculations provided by the SB-MS decoder in the first stage. Specifically, we use the \textit{a posteriori} LLR values obtained after the SB-MS decoder satisfies the proposed stopping criterion (these are computed in line 8 of Algorithm 3 by adding the CN-to-VN and VN-to-CN messages) as initial values for the edge variables $\bar{u}_{i,j}$ of the SB-LP decoder. This provides an improved initialization of the post-processing decoder. 

The SB-MS decoder, which has a faster convergence than SB-LP in general, allows the decoding process to reach a good starting point, reducing considerably the number of errors; if it encounters issues during this process (as detected by the early stopping criterion), the SB-LP decoder is then invoked to remove the small set of residual errors, which are linked to trapping sets. 

\begin{figure}[t]
\resizebox{\columnwidth}{!}{\includegraphics{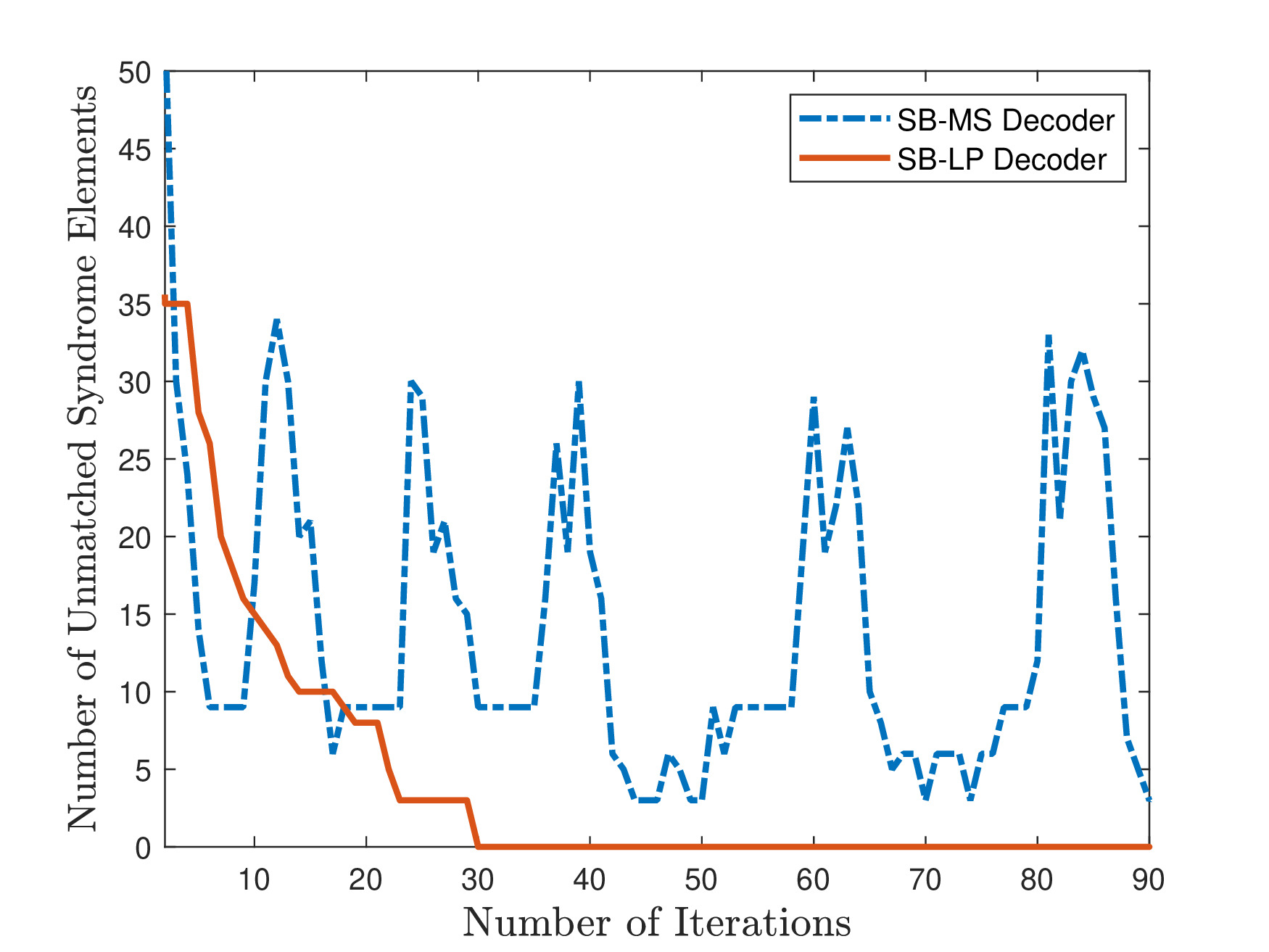}}
\caption{Evolution of the number of unmatched syndrome elements with successive decoding iterations for a sample error vector. Plots are shown for the SB-MS (blue) and SB-LP (red) decoders for the $[[882,24]]$ QLDPC code B1 from \cite{Panteleev_2021}.}
\label{fig:Ocillation}
   \end{figure}
   
\begin{algorithm}
\label{MS+LP}
\caption{Combined SB-MS + SB-LP Decoding Algorithm }
\begin{algorithmic}[1] 
\STATE Input: syndrome $\textbf{s}$
\STATE Initialize parameters:\\
$\lambda_j$ = $\ln \left( \frac{1-(2p/3)}{2p/3} \right)$, $\forall j\in \{1,2,\ldots,n\}$ 
\\
$u_{i,j}=0, \forall  (i,j) \in \mathcal{E}$\\
$I=0$, 
$\hat{\textbf{s}}=\mathbf{0}$
\STATE \textbf{do}
\\
\STATE\hspace{0.4cm}$\hat{\textbf{s}}_{\textrm{pre}}=\hat{\textbf{s}}$ \\
\STATE\hspace{0.4cm}Execute lines (4-11) of Algorithm 1 \\
\STATE\hspace{0.4cm}$I=I+1$\\
\STATE \textbf{while}  ${d_\textrm{H}}(\hat{\textbf{s}}_{\textrm{pre}},\hat{\textbf{s}} )$ $>$ $d_v$ and $I$ $\leq$ $I^{\textrm{(MS)}}_{\textrm{max}}$ 
 \STATE  $\overline{u}_{i,j}=u_{i,j}+v_{i,j} , \forall (i,j) \in \mathcal{E}$\\
\STATE$I=0$ 
\STATE  Execute lines (3-15) of Algorithm 2
\end{algorithmic}
\end{algorithm}

\subsection{Early Stopping Criterion for the SB-MS Decoder}

As explained in the previous subsection, an appropriately designed early stopping criterion can reduce the number of decoding iterations by exiting the SB-MS decoder when it becomes trapped and at the same time provide an improved initialization to the SB-LP decoder. Fig. \ref{fig:Ocillation} shows an example of the evolution of the number of unmatched syndrome elements during SB-MS decoding in the case where the decoder becomes stuck in a trapping set. An oscillatory pattern is observed, indicating that the SB-MS decoder is unable to converge to an error pattern that matches the measured syndrome. In contrast, the proposed SB-LP decoder can converge in such cases; however, a high number of iterations is required. 

From Fig. \ref{fig:Ocillation} we can conclude that operating the SB-MS decoder for a fixed number of iterations is not an efficient approach; instead, the SB-MS decoding procedure should terminate when such oscillatory behavior is detected. In Algorithm 3, we terminate the SB-MS decoding when the Hamming distance ${d_\textrm{H}}(\hat{\textbf{s}}, \hat{\textbf{s}}_{\textrm{pre}})$ between the current syndrome estimate $\hat{\textbf{s}}$ and the previous syndrome estimate $\hat{\textbf{s}}_{\textrm{pre}}$ is less than or equal to the VN degree $d_v$; this approach was found to be more robust that attempting to locate the first local minimum of the number of unmatched syndrome elements. By providing the SB-LP decoder with an improved initialization, this early stopping criterion leads to fewer required decoding iterations while also improving the logical error rate.



It is worth mentioning that this early stopping criterion can be applied when using SB-MS in conjunction with any other post-processing technique, such as OSD or stabilizer inactivation, to activate the post-processing exactly when required and reduce the total number of decoding iterations.



\section{Simulation Results and Discussion}

\begin{figure}[t]
\resizebox{\columnwidth}{!}{\includegraphics{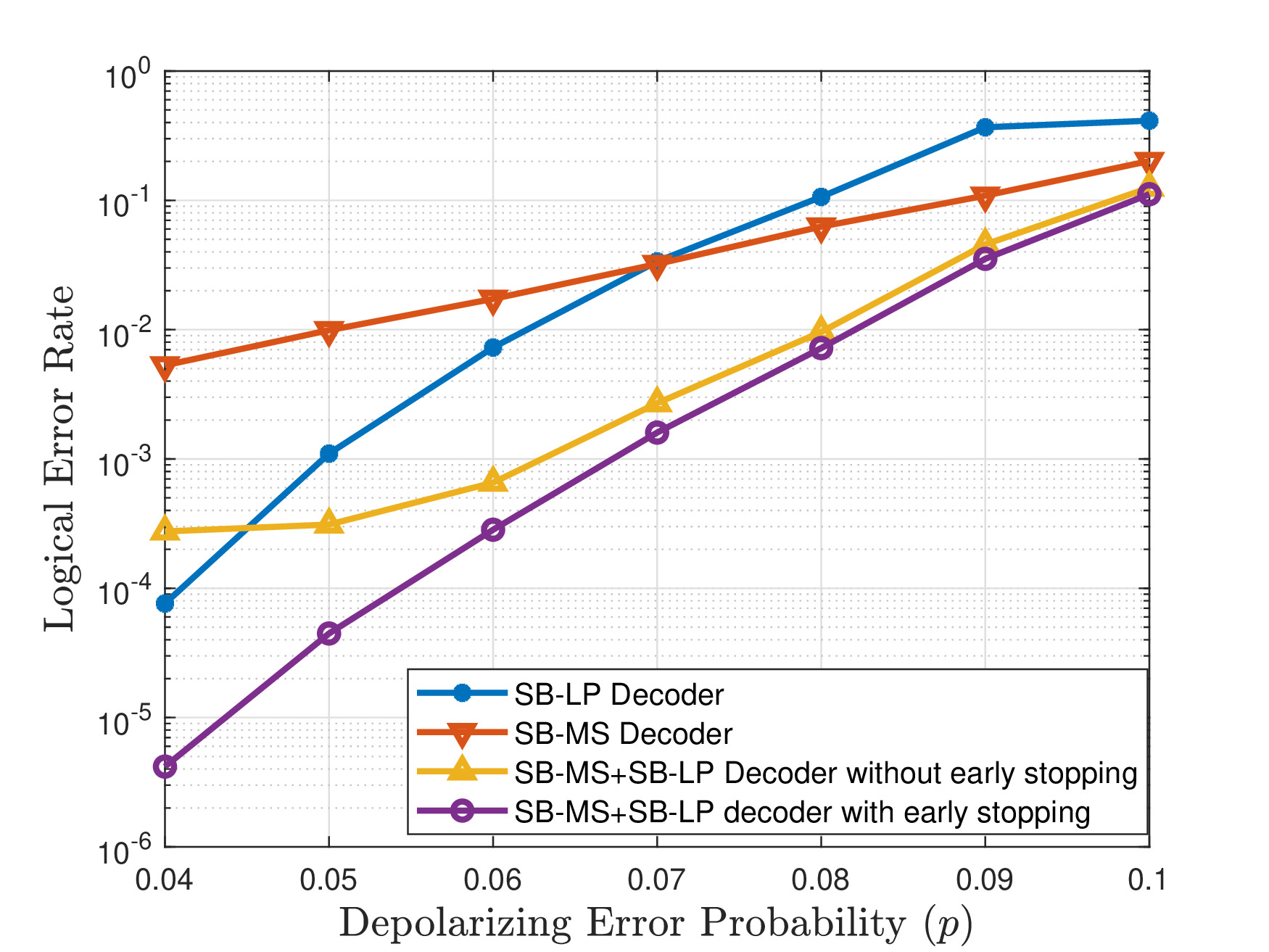}}
\caption{Performance of the SB-MS decoder, the SB-LP decoder, and the combined SB-MS and SB-LP decoder with and without early stopping criterion, for the $[[882,24]]$ QLDPC code B1 from \cite{Panteleev_2021}.}
\vspace{-1.5em}
\label{fig:ms+lp}
\end{figure}


In this section, we compare the simulated performance of the proposed SB-LP decoder with that of SB-MS, and we also evaluate the performance of the combined SB-MS and SB-LP decoders. Simulations are performed for the $[[882,24]]$ QLDPC code designated as code B1 in \cite{Panteleev_2021} (similar results were found for the $[[1922,50]]$ QLDPC code C2 in \cite{Panteleev_2021}; these are omitted due to space limitations). The correction of bit-flip (Pauli $\textrm{X}$) errors is considered under the i.i.d. symmetric depolarizing noise model. 

Fig. \ref{fig:ms+lp} shows the performance curves for the different decoders, with a maximum number of iterations $I^{\textrm{(MS)}}_{\textrm{max}}= I^{\textrm{(LP)}}_{\textrm{max}}=100$ for the SB-MS and SB-LP algorithms respectively; the corresponding scaling factors are $\alpha=0.75$ and $\alpha_1=0.9$. For the case where the SB-MS and SB-LP decoders are combined (Algorithm 3), the first decoder has $I^{\textrm{(MS)}}_{\textrm{max}}=25$ iterations while the second decoder is configured with $I^{\textrm{(LP)}}_{\textrm{max}}=75$. For each point in the figure, $10,000$ logical errors were simulated in order to ensure that the results are statistically significant.

The results reveal that the SB-MS decoder introduces an error floor at a logical error of $10^{-2}$, and is significantly outperformed by the proposed SB-LP decoder. While the SB-LP decoder exhibits a worse threshold behavior, it improves the logical error rate for any depolarizing error probability lower than $0.07$, providing almost two orders of magnitude of improvement in the logical error rate at a depolarizing error probability of $0.04$.

The figure also shows the performance of the combination of the SB-MS and SB-LP algorithms, both with and without the early stopping criterion for SB-MS. Both solutions achieve a good logical error rate for high depolarizing error probabilities (greater than $0.06$), but without the application of the early stopping criterion an error floor is introduced due to the poorer quality of the information passed by the SB-MS decoder to the SB-LP decoder.
\begin{figure}[t]
\resizebox{\columnwidth}{!}{\includegraphics{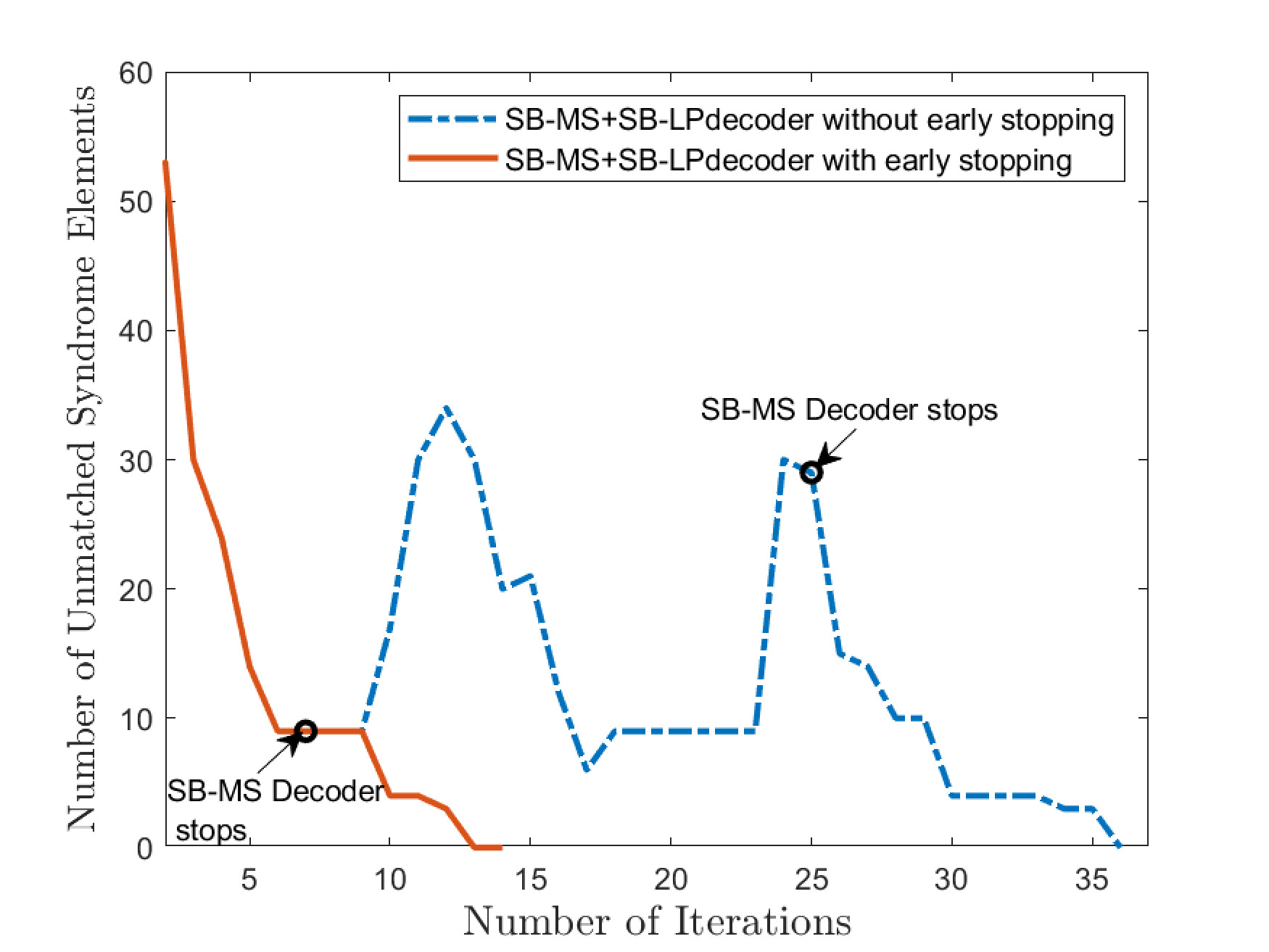}}
\caption{Convergence behavior of the combined SB-MS and SB-LP decoder, with and without early stopping criterion, for the same sample error vector as in Fig.\ref{fig:Ocillation}.
\vspace{-1.5em}}


\label{fig:earlystop}
\end{figure}

\begin{figure}[t]
\resizebox{\columnwidth}{!}{\includegraphics{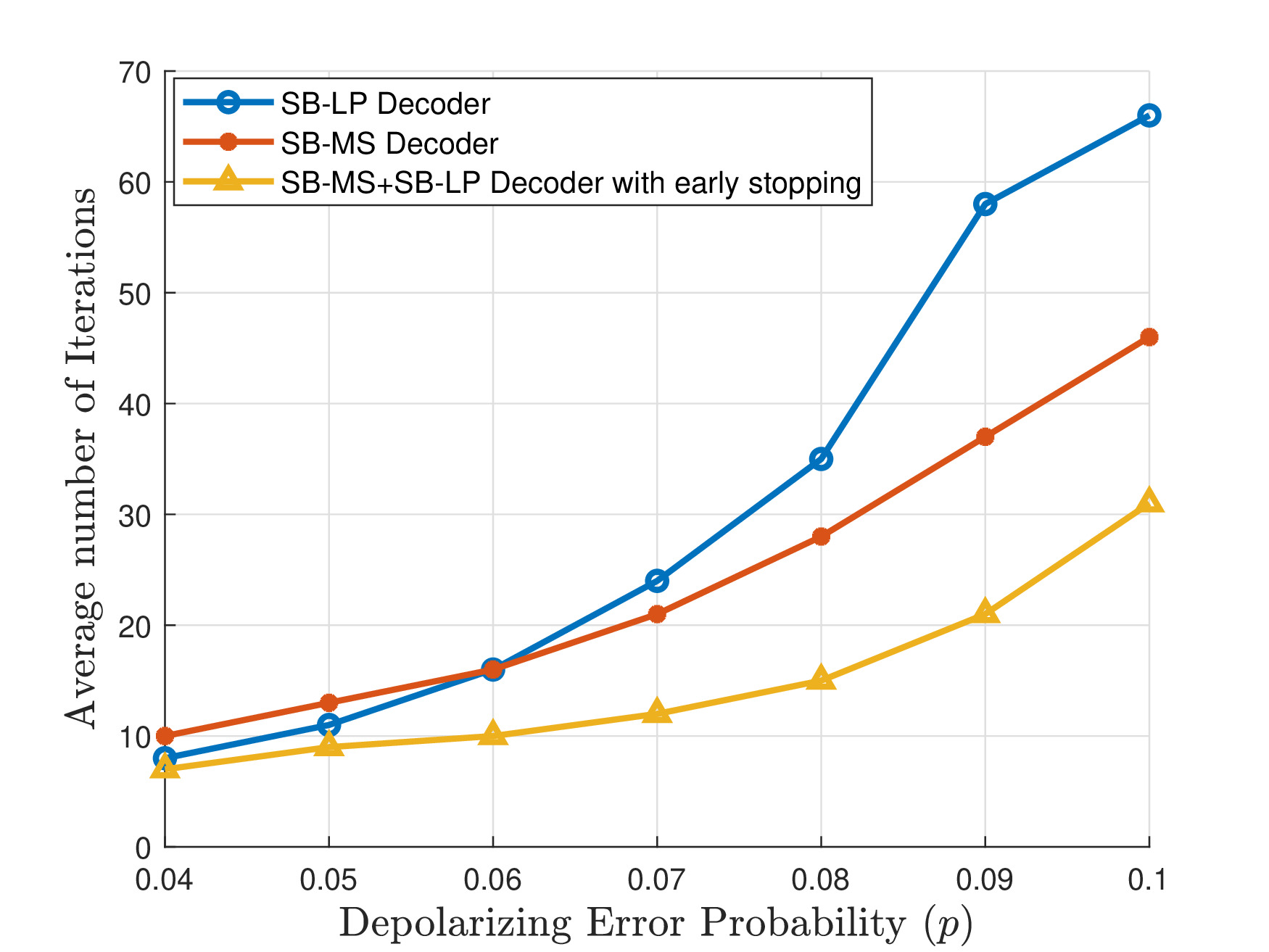}}
\caption{Average number of iterations for the SB-MS decoder, the SB-LP decoder, and the combined SB-MS and SB-LP decoder with and without early stopping criterion, for the $[[882,24]]$ QLDPC code B1 from \cite{Panteleev_2021}.}
\vspace{-1.5em}
\label{fig:ite}
\end{figure}

Fig. \ref{fig:earlystop} illustrates the effect of the early stopping criterion on the result achieved by the SB-MS algorithm. In the case where the early stopping criterion is not employed (dashed line), the SB-MS algorithm will stop when it reaches the maximum number of iterations, while in the case of early stopping (solid line) it can be observed that the SB-MS algorithm will stop near the first local minimum of the number of unmatched syndrome elements. 

Note that, even if the SB-MS decoder is halted after a fixed number of iterations ($25$ in this example), convergence of the subsequent SB-LP algorithm can still be achieved. However, this comes at the cost of a high overall number of iterations ($36$ in this example); in particular, a high number of SB-LP decoder iterations is required because the initialization provided by the SB-MS algorithm is far from optimal. These additional iterations can be sufficient to prevent the decoder's convergence in some cases, introducing the undesirable error floor in Fig.~\ref{fig:ms+lp}. On the other hand, for the same error pattern, employing the early stopping criterion in SB-MS avoids the subsequent oscillations, reducing the number of iterations and providing the SB-LP decoder with a better initialization. For this example, only $7$ SB-MS iterations are required to meet the stopping criterion, followed by $4$ further iterations of the proposed SB-LP decoder. 

Fig. \ref{fig:ite} shows the average total number of iterations as a function of the depolarizing error probability $p$. It can be observed that the average number of iterations of the proposed combined SB-MS and SB-LP with early stopping criterion is always lower than that obtained with SB-MS or SB-LP alone. From the figure, two important conclusions can be drawn. The first is that for high values of depolarizing error probability, the combined SB-MS and SB-LP with early stopping criterion requires a significantly lower number of iterations for convergence than that of the SB-MS or SB-LP decoder alone (note that at these values of $p$, this combined decoder also provides the best logical error rate, as shown in Fig.~\ref{fig:ms+lp}). The second is that for low values of the depolarizing error probability, while there is not a substantial difference in the total number of iterations of the different decoders, the logical error rate of the combined decoder with early stopping criterion sees a significant improvement as shown in Fig.~\ref{fig:ms+lp}; while approximately $10$ iterations are required for all three decoders, the combined decoder provides more than three orders of magnitude improvement in the logical error rate at $p=0.04$.

\section{Conclusion}
We have proposed a novel low-complexity syndrome-based decoder for QLDPC codes based on linear programming, which can be used either as a standalone decoder or as a post-processing decoding step for SB-MS decoding. We also proposed an early stopping criterion for the SB-MS decoder based on the syndrome Hamming distance that judiciously triggers the transition between the two decoding algorithms. The proposed approach was shown to be capable of providing a significant improvement in the logical error rate performance as well as reducing the total number of decoding iterations. 




\bibliographystyle{IEEEtran}
\bibliography{{reference}}

\end{document}